\documentclass[12pt,letterpaper]{JHEP3}
\usepackage{cite}
\usepackage{epsfig}

\title{Flat space physics from holography}

\author{%
Raphael Bousso\\ 
Center for Theoretical Physics, Department of Physics\\
University of California, Berkeley, CA 94720-7300, U.S.A.\\
{\em and}\\
Lawrence Berkeley National Laboratory, Berkeley, CA 94720-8162, U.S.A.\\ 
E-mail: \email{bousso@lbl.gov}}

\abstract{%
  We point out that aspects of quantum mechanics can be derived from
  the holographic principle, using only a perturbative limit of
  classical general relativity.  In flat space, the covariant entropy
  bound reduces to the Bekenstein bound.  The latter does not contain
  Newton's constant and cannot operate via gravitational backreaction.
  Instead, it is protected by---and in this sense, predicts---the
  Heisenberg uncertainty principle.}

\preprint{\hepth{0402058} \\ UCB-PTH-04/02}

\begin{document}

\section{Introduction}

According to the holographic principle~\cite{Tho93,Sus95,ceb2,rmp},
classical spacetime geometry and its matter content arise from an
underlying quantum gravity theory in such a way that the covariant
entropy bound~\cite{ceb1}---thus far only a well-supported conjecture
relating matter entropy to the area of surfaces---is automatically
satisfied.  To some degree, this expectation was borne out by the
AdS/CFT correspondence~\cite{Mal97}, which provides a
non-perturbative, manifestly holographic definition of string theory
in certain spacetimes~\cite{SusWit98}.

By deriving consequences and implications of the holographic
principle, we may learn how specific laws of physics will arise from
quantum gravity even before we know the underlying theory in detail.
An example of such an implication is the generalized second law of
thermodynamics (GSL)~\cite{Bek72,Bek73,Bek74}: the generalized
covariant bound implies~\cite{FMW} that the total entropy of ordinary
matter systems and black holes will never decrease in any physical
process (assuming the ordinary second law holds), a conjecture widely
believed to be true but difficult to prove by other means.

The purpose of this paper is to expose another such implication.
Thus, we adopt the holographic relation between information and
geometry as our axiomatic starting point:
\begin{equation}
S\leq {\Delta A\over 4 l_{\rm Pl}^2}.
\label{eq-gcb1}
\end{equation}
This is a generalized version~\cite{FMW} of the covariant entropy
bound.  The area difference $\Delta A$ will be defined more carefully
in Sec.~\ref{sec-bounds}. For now we note only that the relation
involves a fundamental unit of length, the Planck length, reflecting
its origin in quantum gravity.  (We set $k_{\rm B}=c=1$.)

The entropy $S$ is generally a measure of information; in quantum
field theory, it is identified with the logarithm of the number of
independent Fock space states compatible with the geometric boundary
conditions.\footnote{We stress this distinction since we do not wish
  to assume that the unified theory underlying Eq.~(\ref{eq-gcb1}) is
  in fact quantum mechanical in nature, but rather to leave open the
  possibility that it may encode information by means other than the
  states of a Hilbert space.} But note that $\hbar$ (the value of the
equal-time commutator of a field and its conjugate momentum) is not
explicitly fixed by our assumption.  It will be {\em derived\/}, using
the Raychaudhuri equation of classical general relativity, which
contains only Newton's constant $G$.

As we review in Sec.~\ref{sec-derive}, the Raychaudhuri equation
allows us to eliminate Newton's constant from Eq.~(\ref{eq-gcb1}) in
regions where gravity is weak.  This yields an important intermediate
result, the Bekenstein bound~\cite{Bek81}, which involves only the
combination $l_{\rm Pl}^2/G$ of physical constants.

In Sec.~\ref{sec-heisenberg} we show that the Bekenstein bound would be
violated if the position and momentum uncertainties of a particle were
allowed to become sufficiently small, i.e., if
\begin{equation}
\delta x \, \delta p \ll l_{\rm Pl}^2/G.
\end{equation}
From this we conclude that physical states in Minkowski space must
obey
\begin{equation}
\delta x \, \delta p \gtrsim l_{\rm Pl}^2/G.
\end{equation}
This inequality is the the Heisenberg uncertainty relation.  Thus, 
Planck's constant emerges as a derived quantity,
expressed in terms of the geometric unit of information, $l_{\rm
  P}^2$, and Newton's constant $G$:
\begin{equation}
\hbar\approx l_{\rm Pl}^2/G.
\label{eq-hb}
\end{equation}

Of course, the relation Eq.~(\ref{eq-hb}) is necessary for the first
law of thermodynamics to be satisfied by the entropy $S=\pi R^2/l_{\rm
  Pl}^2$, temperature $T=\hbar/4\pi R$, and mass $M=R/2G$ of a
Schwarzschild black hole.  Indeed, the calculation of the Hawking
temperature remains the only known semiclassical method for
calibrating the numerical coefficients.  But that computation takes
quantum field theory as a starting point.  Here we argue that the
Planck constant can be obtained directly starting from a principle of
quantum gravity.

How can it be legitimate to use classical general relativity to derive
a key aspect of quantum mechanics?  Einstein's theory offers only an
approximate description of Nature, which will surely be transcended
when the problem of unifying quantum theory and gravity is solved.
Quantum mechanics, however, is often assumed to be immune to this
fate.  This prejudice---implicit as soon as we try to ``quantize
gravity''---is not without merit: It has proven difficult to modify
quantum mechanics sensibly.  More importantly, string theory is a
perfectly quantum mechanical theory which does include gravity.
However, it is not clear how much of string theory has been explored,
and how it may be related to a realistic universe.

Indeed, there are reasons to question the assumption that quantum
mechanics is universal.  It is unclear how to ascribe operational
meaning to quantum mechanical amplitudes in highly dynamical spacetime
regions, because experiments cannot be repeated.  Such difficulties
become exacerbated at spacelike singularities.  For quantum mechanical
evolution to proceed, a time coordinate would have to be singled out
among the directions of spacetime.  It would have to survive the
crunch and live on as a quantum mechanical evolution parameter.  But
near a generic singularity, time breaks down as a geometric object no
less than space does.  Attempts to resolve non-timelike singularities
in string theory~\cite{LiuMoo02a,LiuMoo02b,HorPol02,Law02} have so far
only yielded evidence that this covariant behavior persists.

In the absence of a sufficiently general definition of quantum
mechanical observables, it therefore remains conceivable---and perhaps
plausible---that quantum mechanics is no more fundamental than
classical spacetime, and that they both emerge from a unified
description only in certain limits.  The arguments presented in the
present paper are consistent with this viewpoint.  Note that our
derivation applies only in weakly gravitating regions, leaving open
the possibility that in some backgrounds (e.g., in cosmology or
gravitational collapse) quantum mechanics may not emerge in its
conventional form.

\section{Entropy bounds}
\label{sec-bounds}

In this section we introduce various entropy bounds and the
holographic principle.  We explain why we choose Eq.~(\ref{eq-gcb1})
as our axiomatic starting point.  This is a review section and is not
itself part of the derivation.  Thus, we will permit ourselves to
write the Bekenstein bound in terms of $\hbar$, and to make occasional
use of the relation $l_{\rm Pl}^2 = G\hbar$ for the purpose of
elucidating the properties of the bounds.

\subsection{Covariant bound}

The covariant entropy conjecture~\cite{ceb1} applies to any spacetime
region that is well described by classical general relativity.  It
states that the entropy of matter on any light-sheet $L$ of any
two-dimensional surface $B$ obeys
\begin{equation}
S(L)\leq\frac{A}{4 l_{\rm Pl}^2},
\label{eq-cb}
\end{equation}
where $A$ is the area of $B$.  A light-sheet is a 2+1-dimensional
hypersurface generated by nonexpanding light rays orthogonal to $B$.
A full review is found in Ref.~\cite{rmp}.

\begin{figure}[htb!]
  \hspace{.2\textwidth} \vbox{\epsfxsize=.6\textwidth
  \epsfbox{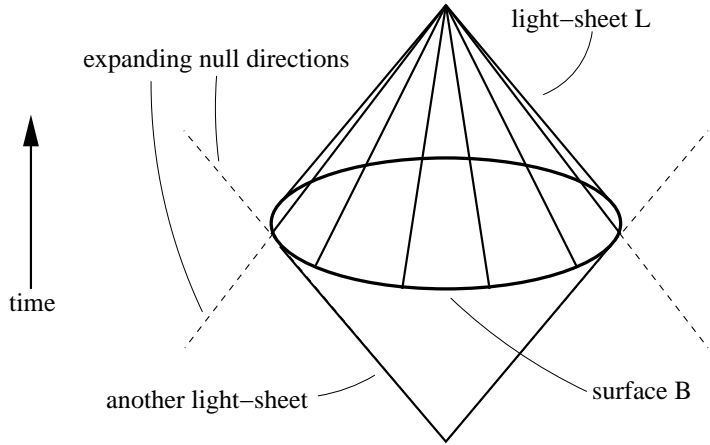}}
\caption%
{\small\sl A two-dimensional surface and its light-sheets}
\label{flekceb}
\end{figure}

Any surface $B$ has four orthogonal null directions, at least two of
which are nonexpanding and give rise to light-sheets.  For example, a
spherical surface in Minkowski space has two light-sheets
corresponding to past and future light-cones ending on $B$
(Fig.~\ref{flekceb}).  When neighboring light-rays intersect, the
expansion becomes positive, and the generating light-rays must be
terminated.  This is why the light-sheets in Fig.~\ref{flekceb} stop
at the tips of the cones.

Gravitational backreaction plays a crucial role in preventing
violations of this bound.  In realistic systems, an increase in
entropy is accompanied by an increase in energy.  Energy focusses
light rays by an amount proportional to $G$.  Thus it hastens the
termination of a light-sheet at caustic points, preventing it from
``seeing'' too much entropy.

We have no quantitative explanation why this backreaction should
always suffice for the holographic bound.  Ultimately this will have
to be explained by an underlying theory in which information, rather
than matter, is fundamental.  But at a qualitative level, the notion
of gravitational focussing appears to capture a key aspect of the
bound.

Decreasing Newton's constant suppresses the gravitational backreaction
and allows more mass and entropy on the light-sheet.  But the bound,
$A/4G\hbar$, is inversely proportional to $G$ and hence compensates by
becoming more lenient as $G\to 0$, and trivial for $G=0$.

\subsection{Bekenstein bound}
\label{sec-bekbound}

A version of Bekenstein's ``universal entropy bound''~\cite{Bek81}
appears as a key intermediate result in our argument.  In its original
form, the bound states that the entropy $S$ of any weakly gravitating
matter system obeys
\begin{equation}
S\leq 2\pi MR/\hbar,
\label{eq-bb}
\end{equation}
where $M$ is the total gravitating mass of the matter and $R$ is the
radius of the smallest sphere that barely fits around the system.  

The absence of Newton's constant in Bekenstein's bound is notable.  It
renders Eq.~(\ref{eq-bb}) independent of the strength of gravity, $G$,
in its regime of validity.  In particular, the bound remains
nontrivial when gravity is turned off completely ($G=0$).  Therefore,
gravitational physics cannot play a role in upholding it, quite unlike
the case of the holographic bound.

If one substitutes for $M$ using the weak gravity condition
\begin{equation}
M\ll R/G,
\label{eq-wg}
\end{equation}
it is immediately apparent that the Bekenstein bound, in its regime of
validity, is significantly tighter than the covariant bound,
Eq.~(\ref{eq-cb}).  To appreciate the difference, consider a single
massive elementary particle.  If we take $M$ to be its rest mass, the
particle can be localized to within a Compton wavelength:
$R\approx\hbar/M$.  Its entropy is of order unity: $S\approx 1$.  The
Bekenstein bound, $2\pi MR/\hbar$, is also of order one, and therefore
is roughly saturated!  The holographic bound, however, is given by the
surface area in Planck units, $\pi R^2/l_{\rm Pl}^2$, which is
typically a huge number.  E.g., for an electron, the holographic bound
gives $S\leq 10^{44}$, which is correct but rather uninteresting.

In recent years the holographic bound, despite its relative weakness,
received far more attention than the Bekenstein bound.
't~Hooft~\cite{Tho93} and Susskind~\cite{Sus95} ascribed fundamental
significance to a bound in terms of area, asserting that the number of
degrees of freedom in quantum gravity is given by the area of surfaces
in Planck units---a conjecture that received strong support by
subsequent developments in string theory~\cite{Mal97}.  Moreover,
their holographic bound has turned out to admit the covariant
formulation (\ref{eq-cb}), whose apparent validity in strongly
gravitating regions provides the most convincing evidence yet for the
significance of the holographic principle in all spacetimes.

The Bekenstein bound has lacked a similar interpretation as a direct
imprint of fundamental physics.  It has been regarded mainly as a
practical limit on information storage and transfer.  Moreover, it
contains quantities (energy, radius) which are well-defined only in
special backgrounds.  This appeared to preclude its generalization to
arbitrary spacetimes---a prerequisite for a fundamental role.  The
result of Ref.~\cite{Bou03}, however, clarifies that the Bekenstein
bound should properly be viewed as expressing the constraints placed
by the holographic principle on the physics of flat space.

\subsection{Generalized covariant bound}

The Bekenstein bound is much tighter than the covariant bound, but
also much less generally applicable.  Thus it is clear that neither
bound can imply the other.  This has obscured the relationship of
Bekenstein's bound to the holographic principle.  

Soon after it was first proposed, it was noticed~\cite{FMW} that the
covariant bound implies the generalized second law of thermodynamics
(GSL) \cite{Bek72,Bek73} for matter that collapses to form a black
hole.  However, in the form Eq.~(\ref{eq-cb}) the covariant bound is
not strong enough to imply the GSL for matter that is added to an
existing black hole.  This motivated Flanagan, Marolf, and
Wald~\cite{FMW} to write down a stronger version of the covariant
bound,\footnote{In Ref.~\cite{FMW} the emphasis was on showing that
certain local conditions on entropy density and energy density are
sufficient for the (generalized) covariant bound (see also
Ref.~\cite{BouFla03}).  Here we make no reference to such
phenomenologically motivated assumptions and derivations.  Rather, we
will consider the GCEB as an axiomatic starting point.}  from which
the GSL does follow in all cases.

\begin{figure}[htb!]
  \hspace{.25\textwidth} \vbox{\epsfxsize=.5\textwidth
  \epsfbox{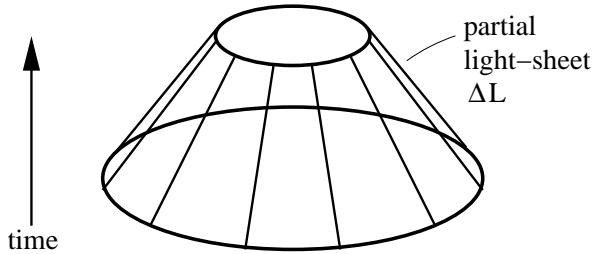}}
\caption%
{\small\sl Partial light-sheet contracting from area $A$ to $A'$.}
\label{flekgceb}
\end{figure}
Consider a partial light-sheet $\Delta L$, which is terminated
prematurely, before the light-rays self-intersect
(Fig.~\ref{flekgceb}).  In general, $\Delta L$ will not capture as
much matter (and as much entropy) as the fully extended light-sheet.
On a partial light-sheet, the final area spanned by the light-rays,
$A'$, will be nonzero and, by the nonexpansion condition, $A'\leq A$.
Thus it is natural to conjecture a ``generalized'' covariant entropy
bound (GCEB),
\begin{equation}
S(\Delta L)\leq\frac{A-A'}{4 l_{\rm Pl}^2}.
\label{eq-gcb}
\end{equation}
When this inequality is applied to weakly gravitating systems, it does
in fact imply the Bekenstein bound~\cite{Bou03}, as we will now show.

\section{Derivation of the Bekenstein bound}
\label{sec-derive}

In this section, we show that the GCEB implies the Bekenstein bound.
Instead of reproducing the rigorous derivation recently given in
Ref.~\cite{Bou03}, we will present a simplified argument which
captures the main idea and yields the Bekenstein bound up to factors
of order unity.

Consider an arbitrary, weakly gravitating system of mass $M$, which
fits into a sphere of radius $R$---for example, the earth.  In order
to apply the covariant bound, one has to construct a light-sheet that
intersects the worldvolume of the earth.  There are many
possibilities, and most will not lead to interesting results.  For
example, one could begin a light-sheet on the surface of the earth and
follow the null geodesics towards the center of the earth.  But then
the final area on the light-sheet would vanish: $A'=0$.  Thus, we
would fail to exploit the power of the tighter bound (\ref{eq-gcb}).
We would learn merely that the entropy is smaller than the surface
area of the earth, which is a far weaker statement than the Bekenstein
bound.

\begin{figure}[htb!]
  \hspace{.05\textwidth} \vbox{\epsfxsize=.9\textwidth
  \epsfbox{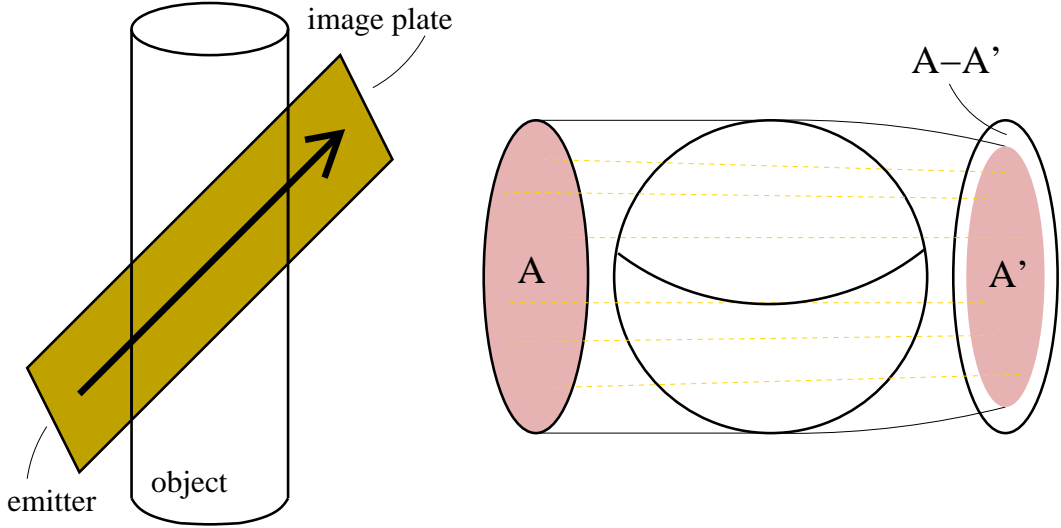}}
\caption%
{\small\sl X-raying a weakly gravitating object.  Left: spacetime
  view; right: spatial view}
\label{flekbend}
\end{figure}

A better strategy is to take advantage of the fact that the covariant
bound does not require the initial surface $A$ to be closed.  This
allows us to ``X-ray'' the system from the side.  Consider parallel
light rays which are emitted from a flat disk\footnote{More precisely,
one must choose a surface on which the initial expansion of the light
rays vanishes exactly.  For weakly gravitating systems, this can
always be arranged by deforming the disk slightly to compensate for
deviations from flat space~\cite{Bou03}.}  tangential to the earth,
with area $A=\pi R^2$.  An image plate of equal area $A$ may be placed
on the opposite side of the earth (Fig.~\ref{flekbend}).

We have arranged for the ``X-rays'' to start out exactly parallel to
each other.  But as they traverse the earth, they will be focussed by
its mass and hence will begin to contract.  Hence, the final area,
$A'$, will be slightly smaller than $A$.  The rays will not illuminate
all of the final plate, but will miss an annulus of area $A-A'$.
Because gravity is weak, this area is determined entirely by the
contraction of the outermost light rays (the ones that just skim the
earth's surface).  It can be quickly estimated from the standard
bending of light effect.

The deflection angle is of order $GM/R$.  Hence, the width of the
annulus is of order $GM$, and the area is
\begin{equation}
A-A'\approx GMR.
\end{equation}
Thus, Eq.~(\ref{eq-gcb}) implies that the entropy of the system obeys
\begin{equation}
S\lesssim \frac{MR}{l_{\rm Pl}^2/G}.
\label{eq-bbl}
\end{equation}
This is the Bekenstein bound (up to factors of order one).\footnote{
With the relation $l_{\rm Pl}^2=G\hbar$, we could transform this
result to the standard form (\ref{eq-bb}).  This would be appropriate
if we were interested in testing the GCEB through tests of the
Bekenstein bound, as in Ref.~\cite{Bou03a}.  Here, however, we take
the GCEB to be true axiomatically and aim to derive Eq.~(\ref{eq-hb})
from it.  Hence we shall retain the form (\ref{eq-bbl}).}  The full
derivation~\cite{Bou03} is independent of any assumptions about the
shape of the object and the distribution of stress-energy.  The
agreement with the Bekenstein bound then includes the numerical
prefactor.\footnote{For non-spherical systems, one finds that the
result obtained from the GCEB is actually stronger than
(\ref{eq-bb}).}

It should be stressed that our understanding of the Bekenstein bound
is still imperfect.  The bound is quite sensitive to the precise
definition of the entropy contained in a finite region.  As entropy is
fundamentally a nonlocal concept, such definitions are fraught with
difficulties.  The derivation we have presented does not automatically
resolve this problem, because the entropy $S$ just goes along for the
ride.


It will be especially important to understand whether the bound
applies to any field theory satisfying reasonable energy conditions,
or whether it imposes further restrictions on the Lagrangian.  For
example, theories with an astronomical number of species appear to be
incompatible with the bound in the formulations proposed in
Refs.~\cite{Bou03a,Bou03b}.  By extension, these questions apply also
to the GCEB.  They are less crucial for the original covariant bound.
The argument in this paper is not sensitive to these issues, but it
does assume that some rigorous formulation of the bounds exists.

\section{Derivation of the uncertainty principle}
\label{sec-heisenberg}

Recall that so far, we have worked entirely in terms of a fundamental
length scale, $l_{\rm Pl}$, which characterizes the maximal
information content of the light-sheets of a given surface as a
function of its area; and Newton's constant, $G$, which measures the
geometric focussing power of the stress-energy of matter.  We have not
made any explicit use of quantum field theory.

To obtain finite microcanonical entropy, as demanded by the entropy
bound, a discretization scheme for physical fields is necessary.  The
canonical equal-time quantization of a relativistic field introduces
$\hbar$ as a constant that defines the commutator between a field
operator and its canonical conjugate.  For example, a scalar field
obeys:  
\begin{equation}
[\phi(\mathbf{ x}), \dot\phi(\mathbf{ x'})]
=i\delta(\mathbf{ x}-\mathbf{ x'})\hbar.
\end{equation}

In the one-particle sector of Fock space, this implies a
nontrivial commutation relation between the position and momentum
operators,
\begin{equation}
[\hat x,\hat p_x]=i\hbar,\mbox{~\it etc.},
\end{equation}
which in turn leads to the Heisenberg uncertainty principle
\begin{equation}
\delta p_x\delta x\geq \hbar/2,\mbox{~\it etc.},
\end{equation}
where $\delta x\equiv\langle(\hat x-\langle \hat
x\rangle)^2\rangle^{1/2}$ and $\delta p_x\equiv\langle(\hat
p_x-\langle \hat p_x\rangle)^2\rangle^{1/2}$.  This inequality can be
saturated by Gaussian wavepackets.

A particle of rest mass $m$ which is at least marginally relativistic
($\mathbf{p}^2\gtrsim m^2$) has a dispersion relation
\begin{equation}
E= \sqrt{m^2+\mathbf{p}^2} \approx |\mathbf{p}|,
\end{equation}
Consider a Gaussian wavepacket with vanishing momentum expectation
value, and with momentum uncertainty $\delta p_x\approx\delta
p_y\approx\delta p_z$.  Its mass is given by
\begin{equation}
M = \langle E\rangle\approx |\delta p_x|.
\end{equation}
Hence, the particle obeys
\begin{equation}
M\delta x\approx\hbar.
\label{eq-jj}
\end{equation}

The spatial size of the particle is approximately its position
uncertainty, $R\approx\delta x$.  The entropy of a typical particle is
given by the logarithm of the number of states of different spin and
hence is of order one.  Even when several different particle species
are allowed, one still obtains $S\approx 1$ with the fields we observe
in Nature.  Hence, Eq.~(\ref{eq-jj}) would violate the bound
(\ref{eq-bbl}) unless
\begin{equation}
\hbar\gtrsim l_{\rm Pl}^2/G.
\label{eq-hbr}
\end{equation}
The right hand side of the Heisenberg uncertainty relation (and of the
canonical commutation relations) is thus determined from the
holographic entropy-area relation and classical
gravity.\footnote{Note that this is also implied by the more rigorous
formulation of the Bekenstein bound recently proposed in
Ref.~\cite{Bou03b}.  There the bound takes the form $S\leq 2\pi^2K
G\hbar/ l_{\rm Pl}^2$, where $K$ labels the Fock space sector of DLCQ.
One immediately obtains $\hbar\gtrsim l_{\rm Pl}^2/G$ by specializing
to $K=1$ and taking the logarithm of the number of particle species to
be of order one ($S_{K=1}\approx 1$).}

To turn the inequality (\ref{eq-hbr}) into an approximate equality, we
invoke economy and assume that $\hbar$ should be chosen such that it
becomes possible to saturate the Bekenstein bound approximately.  The
example of an elementary particle discussed in Sec.~\ref{sec-bekbound}
demonstrates that the bound can indeed be roughly saturated when
$G\hbar/l_{\rm Pl}^2$ is set to unity.

To go further and obtain an exact equality, $\hbar=l_{\rm Pl}^2/G$,
one would need at least one example of a system that precisely
saturates the Bekenstein bound with this value.  This is an important
outstanding problem, along with the question of the proper definition
of entropy in a strict formulation of the bound.

\acknowledgments

I would like to thank M.~Aganagic, T.~Banks, J.~Bekenstein,
E.~Flanagan, D.~Marolf, J.~Polchinski and L.~Susskind for discussions.

\bibliographystyle{board}
\bibliography{all}
\end{document}